\documentclass[amsmath,amssymb,showkeys,superscriptaddress,aps,prd,10pt,twocolumn]{revtex4-2}
\usepackage{graphicx}
\usepackage[caption=false]{subfig}
\usepackage{multirow}
\usepackage{appendix}
\usepackage{graphicx,epstopdf}
\usepackage{subfig}
\def\pslash{p\!\!\!\slash }
\def\qslash{q\!\!\!\slash }
\def\xslash{x\!\!\!\slash }
\def\eslash{\varepsilon\!\!\!\slash }

\usepackage{colortbl}
\usepackage{xcolor}
\usepackage[colorlinks=true, urlcolor=blue, linkcolor=blue, citecolor=blue]{hyperref}
\usepackage{float}

\begin{document}

\title{Magnetic moments of spin--1/2 triply-heavy baryons: A study of Light-cone QCD and Quark-diquark model}

\author{Halil Mutuk}
 \email{hmutuk@omu.edu.tr}
 \affiliation{Department of Physics, Faculty of Arts and Sciences, Ondokuz Mayis University, Atakum, 55200 Samsun, Turkey}
\author{Ulaş Özdem}
\email[]{ulasozdem@aydin.edu.tr}
\affiliation{ Health Services Vocational School of Higher Education, Istanbul Aydin University, Sefakoy-Kucukcekmece, 34295 Istanbul, Turkey}

\begin{abstract}
In this study, the magnetic moments of the spin-1/2 triply-heavy baryons have been calculated using both light-cone QCD sum rules and Quark-diquark model. Theoretical investigations on magnetic  moments of the triply-heavy baryons, are crucial as their results can help us better understand their internal structure and the dynamics of the QCD as the theory of the strong interaction. We compare the results extracted for the magnetic moment with the existing theoretical predictions. It is seen that the obtained magnetic moment values are quite compatible with the results in the literature.
\end{abstract}

%\pacs{Later}% PACS, the Physics and Astronomy
\maketitle
\section{Introduction}
The theoretical and experimental studies of heavy baryons have become a major topic in the literature since properties of heavy baryons provide valuable insight into the nonperturbative aspects of QCD. Quark model predicts charmed or bottomed single, doubly and triply heavy baryons with spin-1/2 and spin-3/2. Among these heavy baryons, triply heavy baryons which sit at the uppermost layer of SU(3) flavor symmetry, may open a door for the understanding of the strong interaction between heavy quarks in the absence of any light quarks. Charm and beauty quarks except top quark which cannot make bound state, are much heavier then the rest of the quarks of the Standard Model, so baryons with three-heavy quarks will refrain light-quark contaminations through the dynamics of strong interactions and can help us better to understand the triply-heavy baryons.

In recent years, there are a lot of works devoted to the singly-heavy and doubly-heavy baryon in the literature. Up to now, all baryons containing a single charm quark have been observed as predicted by the quark model. The experimental progresses on the spectroscopy of the heavy baryons have stimulated the theoretical studies. For example, the observation of doubly-charmed baryon state $\Xi_{cc}^{++}$ by the LHCb collaboration  triggered many new studies \cite{LHCb:2017iph}. This state fills well in the quark model as the $ccu$ baryon. Especially, very recent observation of a narrow resonance structure $X(6900)$ and a broad structure just above the $J/\psi J/\psi$ mass with global significances of more than 5$\sigma$ may open a new era on the fully-heavy exotic multiquark states \cite{LHCb:2020bwg}. 

By this time, no triply-heavy baryon state is reported experimentally. The production of triply-heavy baryons is extremely difficult. This is because one needs to produce three heavy quark-antiquark pairs in one collision event, and the three heavy quarks thus produced need to be close enough to each other in coordinate and momentum space
to enable hadronization into a fully heavy baryon. Nonetheless, there are studies related to the production mechanisms of triply-heavy baryons. %In \cite{Baranov:2004er},  the total and differential cross sections for the production of triply charmed baryons were calculated and it was found that triply charmed baryons may not be observed in $e^+e^-$ collisions. 
In \cite{Baranov:2004er},  the total and differential cross sections for the production of $\Omega_{ccc}$ baryon have been calculated. The predicted value of cross-section of $\Omega_{ccc}$ baryon is close to that the cross-section for $\Omega_{scb}$ baryon production in $e^+e^-$ collisions if the
s-quark mass is set to $300$~MeV. Cross-section calculations of triply-heavy baryons at LHC were done in \cite{Saleev:1999ti,GomshiNobary:2003sf,GomshiNobary:2004mq,GomshiNobary:2005ur}. In \cite{Chen:2011mb}, hadronic production cross-sections of the baryons with the  $ccc$ and $ccb$ quark content were estimated with a conclusion that around 10$^4$-10$^5$ events of triply-heavy baryons can be accumulated for 10 fb$^{-1}$ integrated luminosity at LHC. The production of $ccc$ baryon by quark coalescence mechanism
in high energy nuclear collisions was studied in \cite{He:2014tga} and nuclear medium effect on multicharmed baryon production was studied in \cite{Zhao:2017gpq}. Both of these two studies presented that it is most probable to observe triply-heavy baryon in heavy ion collisions. It was pointed out in \cite{Flynn:2011gf,Wang:2018utj} that it will be good to look for triply-heavy baryons in their semi-leptonic and non-leptonic decays. A recent review for exotic hadrons from heavy ion collisions can be found in \cite{ExHIC:2017smd}. 
Apart from production mechanisms, the spectrum of the triply-heavy baryon states
have been studied extensively via different theoretical approaches like lattice QCD \cite{Meinel:2012qz,Briceno:2012wt,PACS-CS:2013vie,Alexandrou:2014sha,Padmanath:2013zfa,Brown:2014ena,Can:2015exa,Mathur:2018epb}, QCD Sum Rules \cite{Zhang:2009re,Wang:2011ae,Aliev:2012tt,Aliev:2014lxa,Wang:2020avt,Alomayrah:2020qyw,Wu:2021tzo}, potential models \cite{Hasenfratz:1980ka,Silvestre-Brac:1996myf,Migura:2006ep,Jia:2006gw,Roberts:2007ni,Martynenko:2007je,Patel:2008mv,Bernotas:2008bu,Llanes-Estrada:2011gwu,Vijande:2015faa,Thakkar:2016sog,Shah:2017jkr,Shah:2018div,Shah:2018bnr,Liu:2019vtx,Yang:2019lsg,Gordillo:2020sgc}, Faddeev equation \cite{Radin:2014yna,Qin:2019hgk,Yin:2019bxe,Gutierrez-Guerrero:2019uwa} and Regge trajectories \cite{Wei:2015gsa,Wei:2016jyk}.

The  electromagnetic  properties  like  masses and decay widths are  the  important  parameters  of  a  hadron,  which are  measurable  and  computable. Among these electromagnetic properties, the study of magnetic moments of hadrons is an important effort since magnetic moments give information on the internal structure of hadrons. 
The present paper focuses on the magnetic moments of the triply-heavy baryons in light-cone QCD sum rules (LCSR) and quark-diquark model (QDM). 
The motivation of both using LCSR and QDM is to compare the results and make reliable estimations. We believe that the predictions of different methods can be useful for future theoretical and experimental studies. Both LCSR and QDM models are nonperturbative methods and the underlying interactions and assumptions (if exist) can be different. If different models predict same or comparable results, this will pave the way for the true understanding of the related problem. 
In the LCSR method, the operator product expansion is performed  over a twist near the light cone, $x^2 \sim0$ and the nonperturbative contributions show up in the matrix elements of nonlocal operators, which are described  with respect to the light-cone distribution amplitudes (DAs) of the photon~\cite{Chernyak:1990ag, Braun:1988qv, Balitsky:1989ry}, in place of  the vacuum condensates that show up in the traditional QCD sum rules approach \cite{Shifman:1978bx}. Quark model has often been employed  for exploratory studies in QCD. For example, definition of an exotic state stems from the quark model. Any deviation from the quark model predictions of their masses, decay widths, various reactions, production and decay behaviors may also provide insightful clues in the search of the exotic states \cite{Chen:2016qju}. The concept of diquark as an effective degrees-of-freedom is very helpful within the quark model phenomenological approach to simplify the description of hadrons; conventional or exotic.

The paper is organized as follows: in Section \ref{sec:level2}, the methodology of the LCSR and QDM is given. In section \ref{sec:level3}, we present our magnetic moment results with a comparison of references from the literature. 
Section \ref{sec:level4} includes our concluding remarks. The explicit expressions of the magnetic moment of the triply-heavy baryons in LCSR is presented in Appendix.

\section{\label{sec:level2}Methodology}
Different models have been proposed to describe the structure of baryons. In this work, we have used two approaches: LCSR and QDM. Both approaches heave their characteristic features. For example, QCD sum rule is a semi-phenomenological framework to extract spectroscopic information from the QCD lagrangian. This model allows one to relate the hadron spectrum to the fundamental QCD lagrangian. QDM have been applied  for doubly-heavy and triply-heavy baryons since it reduces degrees of freedom and makes calculations easier. In the next two subsections, descriptions of LCSR and QDM will be given.

\begin{widetext}
\subsection{Light-cone QCD sum rules}\label{subLCSR}

To get the magnetic moments of triply-heavy baryons we begin our analysis by considering the following the correlation function
\begin{equation}
 \label{edmn01}
\Pi(p,q)=i\int d^{4}xe^{ip\cdot x}\langle 0|\mathcal{T}\{J_{B_{QQQ'}}(x)
\bar J_{B_{QQQ'}}(0)\}|0\rangle_{\gamma},
\end{equation}%
where the subindex $\gamma$ stands for the background electromagnetic field and $J_{B_{QQQ'}}$ is the interpolating current of the triply-heavy baryons. The most general form of the interpolating current for the $J^P =\frac{1}{2}^+$ baryons under investigation can be written as  
\begin{equation}
\label{cur1}
  J_{B_{QQQ'}}(x) = 2 \varepsilon^{abc} \Big\{ \big(Q_a^T(x)C Q'_b(x)\big)\gamma_5 Q_c(x)
 +t \big[\big( Q_a^T(x)C\gamma_5 Q'_b(x)\big) Q_c(x)  \big]\Big\},
\end{equation}
where $Q$, $Q'$ = b and c-quark, the $a$, $b$ and $c$ are color indexes, C is the charge conjugation and $t$ is arbitrary mixing parameter.

The correlation function in Eq. (\ref{edmn01}) can be obtained in two different representations. On the hadronic representation it is calculated with respect to the hadronic observables while the QCD representation it is calculated with respect to the quark-gluon degrees of freedom. Equating these two different representations then gives us LCSR for physical parameters under investigation. To eliminate the contributions coming from the higher states and continuum  we perform Borel transformation and continuum subtraction to both representations of the obtained LCSR.

For $p^2> 0$, $(p+q)^2 > 0$, the correlation function can be calculating in terms of hadronic parameter.
We insert a complete set of intermediate triply-heavy baryon states  into the correlation function to get the hadronic representation. As a result, we obtain 

\begin{eqnarray}\label{edmn02}
\Pi^{Had}(p,q)&=&\frac{\langle0\mid J_{B_{QQQ'}}\mid
{B_{QQQ'}}(p)\rangle}{[p^{2}-m_{{B_{QQQ'}}}^{2}]}\langle {B_{QQQ'}}(p)\mid
{B_{QQQ'}}(p+q)\rangle_\gamma\frac{\langle {B_{QQQ'}}(p+q)\mid
\bar{J}_{B_{QQQ'}}\mid 0\rangle}{[(p+q)^{2}-m_{{B_{QQQ'}}}^{2}]}+...,
\end{eqnarray}
where $q$ is the photon momentum. 
The matrix elements in the Eq. (\ref{edmn02}) are given as

\begin{eqnarray}\label{edmn03}
\langle0\mid J_{B_{QQQ'}}(0)\mid {B_{QQQ'}}(p,s)\rangle&=&\lambda_{{B_{QQQ'}}}u(p,s),\\
%\end{eqnarray}
%\begin{eqnarray}\label{edmn04}
\langle {B_{QQQ'}}(p)\mid {B_{QQQ'}}(p+q)\rangle_\gamma &=&\varepsilon^\mu\bar u(p)\Bigg[\big(f_1(q^2)+f_2(q^2)\big) \gamma_\mu
+f_2(q^2)\, \frac{(2p+q)_\mu}{2 m_{B_{QQQ'}}}\Bigg]u(p),\label{edmn04}
\end{eqnarray}
where $ u(p)$ and $\lambda_{{B_{QQQ'}}}$ are the Dirac spinor and residue, respectively.
The $f_1(q^2)$ and $f_2(q^2)$ are Lorentz invariant form factors.

Substituting Eqs. (\ref{edmn02})-(\ref{edmn04}) in Eq. (\ref{edmn01}) for hadronic representation we obtain

\begin{align}
\label{edmn05}
\Pi^{Had}(p,q)=&\frac{\lambda^2_{B_{QQQ'}}}{[(p+q)^2-m^2_{B_{QQQ'}}][p^2-m^2_{B_{QQQ'}}]}
  \bigg[\Big(f_1(q^2)+f_2(q^2)\Big)\Big(2\,\pslash\eslash\pslash+\pslash\eslash\qslash+m_{B_{QQQ'}}\,\pslash\eslash
  +
  2\,m_{B_{QQQ'}}\,\eslash\pslash\nonumber\\
  &+m_{B_{QQQ'}}\,\eslash\qslash+m_{B_{QQQ'}}^2\eslash
  \Big)+ other~ structures~proportional~ with~ the~ f_2(q^2) \bigg].
\end{align}

At $q^2=0$, the magnetic moment is described with respect to $f_1(q^2=0)$ and $f_2(q^2=0)$ form factors in the following way;
\begin{equation}
\label{edmn06}
 \mu_{B_{QQQ'}}= f_1(q^2=0)+f_2(q^2=0).
\end{equation}
In this study, we choose the structure $\eslash \qslash$ for analysis, the  triply-heavy baryons have no contaminations. As a result, the hadronic representation of the correlation can be written with respect to magnetic  moment of the spin-$\frac{1}{2}$ triply-heavy baryons as,

\begin{eqnarray}
\label{edmn07}
\Pi^{Had}(p,q)&=&\mu_{B_{QQQ'}}\frac{\lambda^2_{B_{QQQ'}}\,m_{B_{QQQ'}}}{[(p+q)^2-m^2_{B_{QQQ'}}][p^2-m^2_{B_{QQQ'}}]}.
\end{eqnarray}
In obtaining the above expression, summation over spins of $B_{QQQ'}$

\begin{equation}
\label{edmn004}
 \sum_s u(p,s)\bar u(u,s)=\pslash+m_{B_{QQQ'}},
\end{equation}
have also been used. 

In QCD representation, the correlation function in Eq. (\ref{edmn01}), is evaluated in deep Euclidean region with respect to the  QCD degrees of freedom.
For $p^2 <<0$ and $(p+q)^2 <<0$, the main contribution to the correlation function is from small times and small distances~\cite{Colangelo:2000dp}. Hence, the correlation function can be calculated using operator product expansion (OPE). 
To obtain correlation function via QCD degrees of freedom, we insert the explicit forms of the interpolating current in the correlation function and contract the corresponding quark fields with the help of the Wick's theorem. After some simple computation for the correlation function, we get

\begin{align}
\label{edmn08}
 \Pi^{QCD}(p,q)&=4i\, \varepsilon^{abc}\varepsilon^{a^{\prime}b^{\prime}c^{\prime}} \int d^4\,x e^{ip\cdot x}
 \langle 0\mid \Big\{
 \gamma_5 S_Q^{cc^{\prime}}(x)\gamma_5 Tr[\tilde S_Q^{aa^{\prime}}(x)S_{Q'}^{bb^{\prime}}(x)] \nonumber\\
 &
 -\gamma_5 S_Q^{ca^{\prime}}(x)\tilde S_{Q'}^{bb^{\prime}}(x)S_Q^{ac^{\prime}}(x)\gamma_5+t \Big(\gamma_5 S_Q^{cc^{\prime}}(x) Tr[\gamma_5 \tilde S_Q^{aa^{\prime}}(x)S_{Q'}^{bb^{\prime}}(x)]\nonumber\\
 &
 -\gamma_5 S_Q^{ca^{\prime}}(x)\gamma_5\tilde S_{Q'}^{bb^{\prime}}(x)S_Q^{ac^{\prime}}(x)
 +S_Q^{ac^{\prime}}(x)\gamma_5 Tr[\tilde S_Q^{ca^{\prime}}(x) \gamma_5 S_{Q'}^{bb^{\prime}}(x)]\nonumber\\
& -S_Q^{ac^{\prime}}(x) \tilde S_{Q'}^{bb^{\prime}}(x) \gamma_5 S_Q^{ca^{\prime}}(x) \gamma_5 \Big)
+t^2 \Big( S_Q^{cc^{\prime}}(x) Tr[\gamma_5 \tilde S_Q^{aa^{\prime}}(x) \gamma_5 S_{Q'}^{bb^{\prime}}(x) ]\nonumber\\
&
-S_Q^{ca^{\prime}}(x) \gamma_5 \tilde S_{Q'}^{bb^{\prime}}(x)\gamma_5 S_Q^{ac^{\prime}}(x)\Big)
\Big\}\mid 0\rangle_\gamma
\end{align}
where $\tilde S_{Q}^{ij}(x) = CS_{Q}^{{ij}^T}(x)C$ and,$S_Q^{ij}(x)$ is the  heavy quark propagator and it is given as 

\begin{eqnarray}
\label{edmn10}
S_{Q}(x)&=&S^{free}
-\frac{g_{s}m_{Q}}{16\pi ^{2}} \int_{0}^{1}dv~G^{\mu \nu }(vx)\Bigg[ \big(\sigma _{\mu \nu }{\xslash}
  +{\xslash}\sigma _{\mu \nu }\big)\frac{K_{1}( m_{Q}\sqrt{-x^{2}}) }{\sqrt{-x^{2}}}
  %\nonumber\\
%&&
+2\sigma ^{\mu \nu }K_{0}( m_{Q}\sqrt{-x^{2}})\Bigg],
\end{eqnarray}

where 
\begin{align}
S^{free}= \frac{m_{Q}^{2}}{4 \pi^{2}} \Bigg[ \frac{K_{1}(m_{Q}\sqrt{-x^{2}}) }{\sqrt{-x^{2}}}
+i\frac{{\xslash}~K_{2}( m_{Q}\sqrt{-x^{2}})}
{(\sqrt{-x^{2}})^{2}}\Bigg],
\end{align}
with
$K_{i}$'s are Bessel functions of the second kind and $G^{\mu \nu }$ is the gluon field strength tensor.

When the magnetic moment calculation of light quark-containing hadrons is performed using the LCSR method, the correlation function in Eq.~(\ref{edmn08}) includes different contributions: the photon can be emitted both perturbatively or non-perturbatively. But in our case, since triply-heavy baryons contain no valence light quark, the photon cannot be emitted non-perturbatively. Therefore, our calculation includes only the contributions in which the photon is perturbatively emitted.

The propagator of the quark interacting with the photon perturbatively is replaced by
\begin{align}
\label{free}
S^{free}(x) \rightarrow \int d^4y\, S^{free} (x-y)\,\rlap/{\!A}(y)\, S^{free} (y)\,,
\end{align}
the remaining two heavy quark propagators in Eq. (\ref{edmn08}) are with the full quark propagators including the perturbative and non-perturbative parts. The full perturbative contribution is obtained by carrying out the replacement above-mentioned for the perturbatively interacting quark propagator with the photon and making use of the replacement of the remaining propagators by their free parts. The QCD representation of the correlation function can be obtained by applying the Fourier transform to transfer the computations acquired in the x-space to the momentum space.

Having evaluated the correlation function for both physical and QCD representations we now match the coefficient of the structure $\eslash \qslash$ from both representations and carry out double Borel transformation with respect to $p^2$ and $(p+q)^2$. Continuum subtractions, on the other hand, are made using the quark-hadron duality ansatz. %Finally, we obtain
As a results, we get the LCSR for the magnetic moments as
\begin{eqnarray}
\label{edmn14}
 \mu_{B_{QQQ'}}=\frac{e^{\frac{m^2_{B_{QQQ'}}}{M^2}}}{\lambda^2_{B_{QQQ'}}\, m_{B_{QQQ'}}}\, \Delta^{QCD}.
\end{eqnarray}

The explicit expressions of $\Delta^{QCD}$ function is given in the Appendix.

\subsection{Quark-diquark model}
The notion of diquarks dates back to the early days of the quark model. It can be said that diquarks are almost as old as quarks. It provides an alternative description of baryons as bound states of a quark and diquark. The concept of diquarks as effective degrees-of-freedom in quark models has proven useful in the calculation of baryon spectra, see for an instructive review \cite{Anselmino:1992vg} and recent review \cite{Barabanov:2020jvn} and references therein. It leads a simple two-body structure.

A diquark is defined as a colored bound state of two quarks. By this time, no diquark state have been detected in the experiments. However, experimental lack of observation does not eliminate the hypothesis of diquarks as constituents of baryons. Hadrons can exist only when their total color charge of constituent quarks are zero. Technically, this means that every naturally occuring hadron is a color singlet under the group symmetry SU(3).  A diquark is  composed of two quarks ($qq$) whereas conventional quark-antiquark (also called quarkonium) is composed of ($q\bar{q}$). According to group theory, two quarks can attract one another in the $\bar{3}$ representation of SU(3) color, thus a diquark forms having the same color features as an antiquark. In other words, a quark inside the baryon sees a color $\bar{3}$ set of two quarks which is analogous to the antiquark seen by a quark in a traditional meson. According to the perturbation theory of QCD, the potential between two quarks or a quark-antiquark is approximately Coulombic at short distances. At large distances one can invoke lattice QCD approximation and the potential is approximately a sum of two-body linear potentials, although three-body forces exist in baryon \cite{Carlson:1982xi,Isgur:1977ef}. In one-gluon exchange approximation, the quark-quark potential in a baryon is equal to half to the quark-antiquark potential in a meson. This suggests that the interaction between a quark and a diquark in a baryon can be studied in a similar way as the one between a quark and an antiquark in a meson.  

Since $c$ and $b$ quarks are heavy, nonrelativistic treatment is very plausible. In the present work, we choose Cornell potential to study $ccb$ and $bbc$ baryons. It reads as
\begin{equation}
V(r)=\frac{\kappa \alpha_s}{r} + br, \label{Cornell}
\end{equation}
where $\kappa$ is a color factor and associated with the color
structure of the system, $\alpha_s$ is the fine-structure constant of
QCD, and $b$ is the string tension. The first term (Coulomb term) in Eq. (\ref{Cornell}) arises from the one-gluon exchange (OGE) associated with a Lorentz vector structure and the second term (linear part) is responsible for confinement,
which is usually associated with a Lorentz scalar structure.

In the center-of-mass frame, using spherical coordinates, one can factorize the angular and radial parts of the Schrödinger equation. Let $\mu\equiv m_1m_2/(m_1+m_2)$, where $m_1$ and $m_2$ are the constituent masses of quark 1 and quark 2, respectively. Then radial Schrödinger equation can be written as
\begin{equation}
    \left\{  \frac{1}{2\mu}\left[-\frac{\text{d}^2}{\text{d}r^2}+\frac{L(L+1)}{r^2}\right] +V(r) \right\} \psi(r) = E\psi(r).
    \label{schtrans}
\end{equation}
A spin-spin interaction can be included under the assumption of the Breit--Fermi Hamiltonian for OGE \cite{Lucha:1991vn} as:
\begin{align}
    V_S(r) &= -\frac{2}{3(2\mu)^2}\nabla^2V_V(r)\mathbf{S}_1\cdot\mathbf{S}_2 \nonumber\\
    &= -\frac{2\pi\kappa\alpha_S}{3\mu^2}\delta^3(r)\mathbf{S}_1\cdot\mathbf{S}_2 \,.
    \label{spinspin}
\end{align}
The spin-spin interaction in the zeroth-order (unperturbed) potential can be written by replacing the Dirac delta by a Gaussian function 
\begin{equation}
    V_S(r) = -\frac{2\pi\kappa\alpha_S}{3\mu^2}\left(\frac{\sigma}{\sqrt{\pi}}\right)^3\exp\left(-\sigma^2 r^2\right)\mathbf{S}_1\cdot\mathbf{S}_2,
    \label{spinsmear}
\end{equation}
which introduces a new parameter $\sigma$.

With this new definition, the Schrödinger equation of the form of Eq.~(\ref{schtrans}) can be written in a compact form as follows:
\begin{equation}
    \left[  -\frac{\text{d}^2}{\text{d}r^2} +V_{\text{eff}}(r) \right]\varphi(r) = 2\mu E\varphi(r).
    \label{schtranss}
\end{equation}
Here the effective potential $V_{\text{eff}}(r)$ is given by
\begin{equation}
    V_{\text{eff}}(r)\equiv 2\mu\left[ V(r)+V_S(r) \right] + \frac{L(L+1)}{r^2}
    \label{veff}
\end{equation}
and composed of Cornell potential, spin-spin interaction and orbital excitation. 

Computing baryon masses in the quark-diquark model has two steps in which in the first step the diquark mass is calculated and in the second step masses of the system formed by a  quark interacting with this diquark are calculated. As mentioned before, the energy of a quark-diquark pair is assumed to be the same as the one of a quark-antiquark pair. The nonrelativistic Schrödinger equation with Cornell potential is solved and the mass spectra is obtained using
\begin{equation}
M_B=m_{QQ} + m_{Q} + \left\langle H \right\rangle,
\end{equation}
where $m_{QQ}$ is the mass of heavy diquark, $m_{Q}$ is the mass of heavy quark and $\left\langle H \right\rangle$ is composed of Cornell potential and the spin dependent interactions. 

The SU(3) color symmetry of QCD implies that, combining a quark and an antiquark in the fundamental color representation gives $\vert q \bar{q} \rangle:  \textbf{3}  \bigotimes  \bar{\textbf{3}}= \textbf{1} \bigoplus \textbf{8} $. This representation yields the color factor for the color singlet states as $\kappa=-4/3$ of the quark-antiquark system. When we combine two quarks in the fundamental color representation, it reduces to $\vert q q \rangle:  \textbf{3}  \bigotimes  \textbf{3}= \bar{\textbf{3}} \bigoplus \textbf{6} $, a color antitriplet  $\bar{\textbf{3}}$ and a color sextet $\textbf{6}$. Similarly combining two antiquarks reduces to $\vert \bar{q} \bar{q} \rangle:  \bar{\textbf{3}}  \bigotimes  \bar{\textbf{3}}= \textbf{3} \bigoplus \bar{\textbf{6}} $, a triplet $\textbf{3}$ and antisextet $\bar{\textbf{6}}$.
The antitriplet state has a color factor of $\kappa=-2/3$ which is attractive whereas the sextet state  has a color factor of $\kappa=+1/3$ which is repulsive. Therefore we will only consider diquarks in the antitriplet color state. It should be mentioned that going from the color factor $\kappa=-\frac{4}{3}$ (for a quark-antiquark in the singlet color state) to the color factor $\kappa=-\frac{2}{3}$ (for a quark-quark in the antitriplet color state) is equivalent to introducing a factor of 1/2, which can generalized to be a global factor since it comes from the color structure of the wave function. As mentioned before, it is very common to extend this 1/2 factor to the whole potential describing the quark-quark interaction for studying diquark systems.

The parameters of the model are listed in Table \ref{tab:table1}. The parameters are obtained by fitting a quark-antiquark model to available charmonium and bottomonium meson data \cite{Lundhammar:2020xvw}.
\begin{table}[h]
\caption{\label{tab:table1}Potential model parameters.}
\begin{ruledtabular}
\begin{tabular}{cc}
Parameter&Numerical value\\
\hline
$m_c$ & $1.459 ~ \text{GeV}$ \\
$m_b$ & $4.783 ~ \text{GeV}$ \\
$\alpha_s$ & 0.3714 \\
$b$ & $0.1445~ \text{GeV}^2$ \\
$\sigma$ & $1.5 ~ \text{GeV}$
\end{tabular}
\end{ruledtabular}
\end{table}

Equipped with these arguments, we can obtain magnetic moments of $ccb$ and $cbb$ baryons. The magnetic moment of baryons can be obtained in terms of the spin, charge and effective mass of the bound quarks as
\begin{equation}
\mu_B= \sum_i \langle \phi_{sf} \vert \mu_{i}\vec{\sigma}^i \vert \phi_{sf} \rangle, \label{magnetic}
\end{equation}
where 
\begin{equation}
\mu_i=\frac{e_i}{2m_i^{eff}}.
\end{equation}
Here $e_i$ is the charge, $\sigma^i$ is the spin of the  constituent quark in the baryon, $\phi_{sf}$ is the spin-flavour wave function. The effective mass for the each constituent quark $m_i^{eff}$ can be defined as
\begin{equation}
m_i^{eff}=m_i \left( 1+ \frac{\langle H \rangle}{\sum_i m_i} \right),
\end{equation}
where $\langle H \rangle= E + \langle V_S \rangle$ and $m_i$'s are the respective model quark mass parameters. 

It should be noted here that the magnetic moments of triply-heavy baryons have been studied non-relativistic quark-diquark model in \cite{Thakkar:2016sog}.  They use a Coulomb plus power potential with a power index $\nu$  and rather a different type of spin-spin interaction. In our analysis, we use a Cornell potential with spin-dependent interactions. In addition to this, they use a total Hamiltonian which is composed of a diquark Hamiltonian and quark-diquark Hamiltonian. Accordingly, the quark-diquark model of Ref.~\cite{Thakkar:2016sog} is different than our model.

\end{widetext}

\section{\label{sec:level3}Numerical Results and Discussion}

\subsection{Results of Light-cone QCD sum rules}

Now we are ready to investigate numerically the LCSR acquired in the previous subsection and compute the numerical values of the magnetic moments of the triply-heavy baryons. To do this we take the quark masses as their pole values $m_c =1.67 \pm 0.07$~GeV and $m_b =4.78 \pm 0.06$~GeV.
The numerical value of the residues and masses of these baryons is needed to perform the numerical calculations in LCSR. These parameters were obtained using the two-point QCD sum rules in the Ref. \cite{Aliev:2012tt}. The numerical values of these parameters  depend on the arbitrary mixing parameter $t$. To be consistent, we use the same range of $t$ which is $-0.5 \leq t \leq -1.75$.
%Another set of important input parameters are photon distribution amplitudes of different twists. The whole list of these distribution amplitudes together with the values of the input parameters are given in Ref. \cite{Ball:2002ps}.

%
The LCSR for magnetic moments under study are also functions of two more auxiliary parameters that are included at different
stages of the calculations: the continuum threshold $s_0$ and Borel mass parameter $M^2$. Complying with the standard prescription of LCSR method, the  $s_0$ and  $M^2$ are varied to find the optimal stability interval, in which the perturbative contribution should be larger than the non-perturbative contributions while the pole contribution larger than continuum contribution. Therefore, the LCSR intervals are taken as $s_0 =(144-148)$ GeV$^2$, $M^2 = (12-16)$ GeV$^2$ for $\Omega^{0}_{bbc}$ baryon and  $s_0 =(78-82)$ GeV$^2$, $M^2 = (8-12)$ GeV$^2$ for $\Omega^{+}_{ccb}$ baryon, respectively. It is noted here that the magnetic moment results exhibit less dependent variation on the value $s_0$ when the quark masses are taken at the pole value.

Now that we have determined all the input parameters, we can perform our numerical calculations. We summarized our results as follows
\begin{align}
 &\mu_{\Omega^{+}_{ccb}} =0.61 \pm {0.21}~\mu_N,\\
&\mu_{\Omega^{0}_{bbc}} =-0.19 \pm {0.06}~\mu_N.
\end{align}

The errors in our results reflect the uncertainties in the $M^2$, $s_0$ as well as above-mentioned input parameters. We should also emphasize that the main source of uncertainties is the variations in connection with the $s_0$ and the results weakly depend on the variation of the $M^2$.

\subsection{Results of Quark-diquark model}
The quark-diquark model can be used not only to study spectroscopy but also to structure properties such as magnetic moments. In order to obtain magnetic moments, we first calculate diquark masses. In the second step we evaluate baryon masses to get effective masses of quarks inside the baryon. Then magnetic moments can be calculated by using effective quark masses. The results are presented in Table \ref{tab:table2}.

\begin{table}[h]
\caption{\label{tab:table2}Expressions and results for magnetic moments of spin--1/2 triply-heavy baryons using effective quark masses (in nuclear magneton $\mu_N$).}
\begin{ruledtabular}
\begin{tabular}{ccc}
Particle& Expression &Magnetic moment \\
\hline
$\Omega_{ccb}^+$ & $\frac{4}{3}\mu_c-\frac{1}{3}\mu_b$ & 0.581\\
$\Omega_{cbb}^0$ & $\frac{4}{3}\mu_b-\frac{1}{3}\mu_c$ & -0.227\\
\end{tabular}
\end{ruledtabular}
\end{table}

\subsection{Comparison with literature}
In this subsection, for  completeness, we compare our results with the different models from the literature. The results can be seen in Table \ref{tab:table3}. For the sake of comparison we also give different models namely BD, AL1, AL2, AP1, AP2  potential models \cite{Silvestre-Brac:1996myf}, hypercentral constituent quark model (HCCQM) \cite{Patel:2008mv}, non-relativistic quark-diquark model (NRQDM) \cite{Thakkar:2016sog}, power-law potential \cite{Barik:1983ics}, relativistic logarithmic potential (RLP) \cite{Jena:1986xs}, non-relativistic quark model (NRQM) \cite{Silvestre-Brac:1996tmn}, relativistic three-quark model (RTQM) \cite{Faessler:2006ft}, bag model and non-relativistic quark model (NRQM-I) \cite{Bernotas:2012nz}, effective mass (EMS) and screened charge scheme (SCS) \cite{Dhir:2013nka} and extended bag model (EBM) \cite{Simonis:2018rld}.

\begin{table}[htp]
\caption{\label{tab:table3}Comparison of our magnetic moments  with the results of available studies (in unit of $\mu_N$).}
\begin{ruledtabular}
\begin{tabular}{l|c|c}
Models& $\Omega_{cbb}^0$ &$\Omega_{ccb}^+$ \\
\hline
\hline
BD~\cite{Silvestre-Brac:1996myf}& -0.191& 0.466\\
AL1~\cite{Silvestre-Brac:1996myf}& -0.193& 0.475\\
AL2~\cite{Silvestre-Brac:1996myf}& -0.192& 0.471\\
AP1~\cite{Silvestre-Brac:1996myf}& -0.195& 0.479\\
AP2~\cite{Silvestre-Brac:1996myf}& -0.193& 0.473\\
HCCQM~\cite{Patel:2008mv} &-0.203 &0.502 \\
NRQDM~\cite{Thakkar:2016sog} & -0.223& 0.565 \\
Power law~\cite{Barik:1983ics} & -0.197& 0.476 \\
RLP~\cite{Jena:1986xs} &-0.20 & 0.49\\
NRQM~\cite{Silvestre-Brac:1996tmn} & -0.193& 0.475 \\
RTQM~\cite{Faessler:2006ft} & -0.20& 0.53\\
Bag model~\cite{Bernotas:2012nz} &-0.205 &0.505 \\
NRQM-I~\cite{Bernotas:2012nz} &-0.21 &0.54 \\
EMS~\cite{Dhir:2013nka} & -0.205&0.508 \\
SCS~\cite{Dhir:2013nka} & -0.200& 0.522\\
EBM~\cite{Simonis:2018rld} & -0.187& 0.455\\
This Work (QDM) &-0.227 & 0.581 \\
This Work (LCSR) &$-0.19 \pm{0.06}$  & $0.61 \pm{0.21}$  \\
\end{tabular}
\end{ruledtabular}
\end{table}
It can be seen from Table \ref{tab:table3} that magnetic moment of $\Omega_{cbb}^0$ changes from $-0.19 \pm{0.06}$ to -0.227, whereas magnetic moment of $\Omega_{ccb}^+$ changes from 0.466 to $0.61 \pm{0.21}$ in unit of $\mu_N$, respectively. The results of the references agree well with one another. This can be due to the fact that magnetic moments of triply-heavy baryons are virtually governed by the magnetic moments of heavy quarks. Therefore all the models in the table gave almost similar results. 

\section{\label{sec:level4}Concluding Remarks and Final Notes}
In this work, we have calculated magnetic moment of spin--1/2 triply-heavy baryons, $\Omega_{cbb}^0$ and  $\Omega_{ccb}^+$, via light-cone QCD sum rules and quark-diquark model. A quark-diquark interpolating current was used in QCD sum rules analysis and a non-relativistic approach was welcomed in the quark-diquark model.

Our results for the magnetic moments of both Light-cone QCD sum rules and Quark-diquark methods %(LCSR and QDM) 
agree well within each other. Comparing with the predictions of other theoretical methods, it can be seen that our results also agree with the references.

By this time, no experimental data of magnetic moments are available for the triply-heavy baryons. We hope these results will motivate other theoretical studies in the near future. We also look forward to see experimental data on magnetic moments of triply heavy baryons from future experimental facilities.

\begin{widetext}

\subsection*{Appendix: The explicit expressions of  \texorpdfstring{$\Delta^{QCD} $}{} function}

In this Appendix, we present the explicit expression for the function $\Delta^{QCD} $ obtained from the LCSR in subsection \ref{subLCSR}. It is acquired by selecting the $\eslash \qslash$ structure as follows
 %The explicit expressions of $\Delta^{QCD}$ function is given as follows: 
%
\begin{align}\label{resson}
 \Delta^{QCD}&=\frac {3 (-1 + t)^2} {83886080 \pi^6}\Bigg[
   3 e_ {Q^\prime} m_ {Q^\prime } \Big (I[5, 1, 3] - 3 I[5, 1, 4] + 
       3 I[5, 1, 5] - I[5, 1, 6] - 
       3 (I[5, 2, 3] - 2 I[5, 2, 4] + I[5, 2, 5] \nonumber\\
       &- I[5, 3, 3] + 
          I[5, 3, 4]) - I[5, 4, 3]\Big) \nonumber\\
          &+ 
    e_Q m_ {Q^\prime } \Big (I[5, 2, 2] - 2 I[5, 2, 3] + I[5, 2, 4] - 
       3 I[5, 3, 2] + 4 I[5, 3, 3]  - I[5, 3, 4]+ 3 I[5, 4, 2] - 
       2 I[5, 4, 3] \nonumber\\
       &- I[5, 5, 2]\Big) \nonumber\\
       &- 
    e_ {Q^\prime} m_Q \Big (-I[5, 2, 2] + 2 I[5, 2, 3] - I[5, 2, 4] + 
       3 I[5, 3, 2] - 4 I[5, 3, 3] + I[5, 3, 4] - 3 I[5, 4, 2] + 
       2 I[5, 4, 3] \nonumber\\
       &+ I[5, 5, 2]\Big) \nonumber\\
       &- 
    e_Q m_Q \Big (3072 I[5, 3, 1] + 11 I[5, 3, 2] - 28 I[5, 3, 3] + 
        20 I[5, 3, 4] + 3072 I[5, 4, 1] - 22 I[5, 4, 2] + 
        28 I[5, 4, 3] \nonumber\\
    &+ 3072 I[5, 5, 1] + 11 I[5, 5, 2] + 
        3072 I[5, 6, 1]\Big)\Bigg]\nonumber\\
    %    \end{align}
     %   \begin{align}
        %%%%%%%%%%%%%%%%%%%%%%%%%%%%%%%%%%%%%%%%%%%%%
        &+ 
        \frac {m_Q^2 m_ {Q^\prime }} {20971520 \pi^6}\Bigg[- 
      e_ {Q^\prime} (1 + t)^2\Big (I[4, 1, 2] - 2 I[4, 1, 3] + 
       I[4, 1, 4] - 2 I[4, 2, 2] + 2 I[4, 2, 3] + I[4, 3, 2]\Big) \nonumber\\
       &+ 
    e_Q \Big (256 (1 + t)^2 I[4, 2, 1] + (3 - 2 t + 3 t^2) I[4, 2, 
          2] - 4 I[4, 2, 3] - 4 t^2 I[4, 2, 3] + 256 I[4, 3, 1] + 
        512 t I[4, 3, 1] \nonumber\\
        &+ 256 t^2 I[4, 3, 1] - 3 I[4, 3, 2] + 
        2 t I[4, 3, 2] - 3 t^2 I[4, 3, 2] + 
        256 (1 + t)^2 I[4, 4, 1]\Big)\Bigg]\nonumber\\
        %%%%%%%%%%%%%%%%%%%%%%%%%%%%%%%%%%%%%%%5
        &+\frac {\langle g_s^2 G^2 \rangle } {113246208 \pi^6} (-1 + t) \Bigg[
   14 (-1 + t) e_Q m_ {Q^\prime } \Big (I[3, 1, 2] - 2 I[3, 1, 3] + 
       I[3, 1, 4] - 2 I[3, 2, 2] + 2 I[3, 2, 3] \nonumber\\
       &+ I[3, 3, 2]\Big) \nonumber\\
       &+ 
    3 (-1 + t) e_ {Q^\prime} m_ {Q^\prime } \Big (I[3, 1, 2] - 
       2 I[3, 1, 3] + I[3, 1, 4] - 2 I[3, 2, 2] + 2 I[3, 2, 3] + 
       I[3, 3, 2]\Big)\nonumber\\
       &- (1 + 
       t) e_ {Q^\prime} m_Q \Big (64 I[3, 2, 1] - I[3, 2, 2] + 
       2 I[3, 2, 3] + 64 I[3, 3, 1] + I[3, 3, 2] + 
       64 I[3, 4, 1]\Big) \nonumber\\
       &- (1 + 
       t) e_Q m_Q \Big (832 I[3, 2, 1] + 8 I[3, 2, 2] + 
        5 I[3, 2, 3] + 832 I[3, 3, 1] - 8 I[3, 3, 2] + 
        832 I[3, 4, 1]\Big)\Bigg].
\end{align}
The functions I[n, m, l] is defined as:
\begin{align}
 I[n,m,l]=\int^{s_0}_{\alpha} ds \int_0^1 dv \int_0^1 dw \,\big(\alpha + s\big)^n v^m w^l
\end{align}
where $\alpha = (2m_Q+m_{Q^\prime})^2 $. 

It should be noted that in the expressions given in Eq.~(\ref{resson}), we have given only the terms that make significant contributions to the numerical values of the magnetic moments. Contributions not given here are taken into account in numerical calculations, but for simplicity they are not shown in the text.

\end{widetext}

\bibliography{triply-heavy}
\end{document}